\documentclass[3p,times,twocolumn]{elsarticle}

\usepackage{ecrc}

\volume{00}

\firstpage{1}

\journalname{Nuclear Physics B Proceedings Supplement}

\runauth{Evgueni Goudzovski}

\jid{nuphbp}

\jnltitlelogo{Nuclear Physics B Proceedings Supplement}

\usepackage{amssymb}

\usepackage[figuresright]{rotating}

\begin{document}

\begin{frontmatter}



\dochead{}

\title{Rare Kaon Decay Experiments at CERN}


\author{Evgueni Goudzovski\fnref{label1}}
\fntext[label1]{On behalf of the NA62 collaboration: Birmingham, Bratislava, Bristol, CERN, Dubna, Fairfax, Ferrara, Florence, Frascati, Glasgow, Liverpool, Louvain-La-Neuve, Mainz, Merced, Moscow, Naples, Padua, Perugia, Pisa, Prague, Protvino, Rome La Sapienza, Rome Tor Vergata, San Luis Potos\'i, Sofia, Stanford, Turin}

\address{School of Physics and Astronomy, University of Birmingham\\Edgbaston, B15 2TT, United Kingdom}

\begin{abstract}
The NA62 experiment at the CERN SPS aims to measure the branching ratio of the rare decay $K^+\to\pi^+\nu\bar\nu$ with a relative precision of $\sim10\%$. To achieve that goal, it is designed to be exposed to $1.2\times 10^{13}$ $K^+$ decays in its fiducial volume. The unprecedented $K^+$ flux will lead to record sensitivities to rare and forbidden decays of $K^+$ and $\pi^0$, including those that violate lepton flavour or lepton number conservation. The expected NA62 performances for lepton flavour conservation and lepton universality tests are discussed. Relevant on-going or recently completed measurements from the $K^\pm$ decay data sets collected by earlier kaon experiments at CERN (NA48/2 and NA62-$R_K$) are also presented.
\end{abstract}


\end{frontmatter}

\section{Introduction}

The NA62 experiment at the CERN SPS, which is expected to start data taking in October 2014 and operate until 2017, aims to measure the branching ratio (${\cal B}$) of the very rare flavour changing neutral current decay $K^+\to\pi^+\nu\bar\nu$ with a relative precision of $\sim10\%$. In the Standard Model (SM), this decay proceeds via loop and box diagrams and is characterized by maximum CKM suppression. The SM prediction for the decay rate is ${\cal B}(K^+\to\pi^+\nu\bar\nu) = (7.81\pm0.80)\times10^{-11}$~\cite{br11}. The relatively high precision is in part because the hadronic matrix element can be related to the experimentally well known semileptonic decay rate ($K^+\to\pi^0e^+\nu$). The quoted uncertainty is dominated by the CKM parametric one, which is expected to decrease in the near future.

The smallness of the expected ${\cal B}(K^+\to\pi^+\nu\bar\nu)$ means that a very large number $K^+$ decays are required for the measurement. According to NA62 design, about $1.2\times 10^{13}$ $K^+$ will decay inside its fiducial volume over the data taking. Moreover, since many major $K^+$ decay modes (most notably, $K^+\to\pi^+\pi^0$) result in $\pi^0$ production, NA62 will be exposed to a sample of $\sim 10^{12}$ tagged decays of $\pi^0$ produced in vacuum. The resulting world's largest $K^+$ and $\pi^0$ decay samples will provide an opportunity to search for a range of novel phenomena. In particular, NA62 will be well positioned to improve the precision in the searches for $K^+$ and $\pi^0$ decays violating the conservation of lepton flavour or lepton number, as well as lepton universality tests and searches for heavy neutrinos produced in $K^+$ decays.

\section{Lepton flavour violation in kaon decays}

\begin{table*}
\begin{center}
\caption{Current experimental upper limits and lepton flavour and lepton number violating decay rates of $K^+$ and $\pi^0$, which can be potentially be improved by the NA62 experiment.}
\label{tab:lfv}
\begin{tabular}{lclc}
\hline
Mode & Upper limit (90\% CL) & Experiment & Reference\\
\hline
$K^+\to\pi^+\mu^+e^-$    & $1.3\times 10^{-11}$ & BNL E777/E865 & \cite{e777}\\
$K^+\to\pi^+\mu^-e^+$    & $5.2\times 10^{-10}$ & BNL E865 & \cite{e865}\\
$K^+\to\pi^-\mu^+e^+$    & $5.0\times 10^{-10}$ & BNL E865 & \cite{e865}\\
$K^+\to\pi^-e^+e^+$      & $6.4\times 10^{-10}$ & BNL E865 & \cite{e865}\\
$K^+\to\pi^-\mu^+\mu^+$  & $1.1\times 10^{-9}$     & NA48/2 & \cite{ba11}\\
$K^+\to\mu^-\nu e^+e^+$     & $2.0\times 10^{-8}$  & Geneva-Saclay & \cite{gva-sac}\\
$K^+\to e^-\nu \mu^+\mu^+$  & no data  & & \\
$\pi^0\to\mu^\pm e^\mp$     & $3.6\times 10^{-10}$ & KTeV  & \cite{ktev}\\
\hline
\end{tabular}
\end{center}
\end{table*}

Lepton Flavour (LF) conservation is a cornerstone of the Standard Model of particle physics. Historically, LF conservation tests have been the standard probe
for the non-SM phenomena, and the experimental verifications of this concept have contributed to the formulation of the SM itself. However LF conservation is an accidental rather than a fundamental symmetry of the SM. As a result, it is violated in many extensions of the SM, including supersymmetry (via the slepton mixing matrix), models with heavy neutrinos, extra dimensions, and others.

The discovery of neutrino oscillations has led to a conclusion that the LF symmetry is approximate rather than exact. It also implies that the neutrinos have non-zero masses, which (in addition to constituting a non-SM phenomenon by itself) opens the questions about the possible Majorana nature of the neutrino, as well as the origin of the small (sub-eV) observed neutrino mass scale. The LF violating (LFV) processes occur at unobservably small rates in the SM with massive neutrinos, therefore their experimental observation would provide an unambiguous evidence for new physics. This underpins the experimental interest in LFV phenomena, including searches for the neutrinoless nuclear double $\beta$ decays, LFV decays of hadrons ($K$, $B$) and leptons ($\mu$, $\tau$), and $\mu\to e$  conversion.

In this context, the next-generation high luminosity kaon experiments represent opportunities to push down the limits on LFV phenomena in kaon sector. The
LFV decays of $K^+$ and $\pi^0$ are listed in Table~\ref{tab:lfv}: the NA62 experiment should be able to significantly improve on many, if not all, of these limits.

\boldmath
\section{NA48/2 measurement of the $K^\pm\to\pi^\mp\mu^\mp\mu^\mp$ decay}
\unboldmath

Lepton number violation in the $K^\pm\to\pi^\mp\ell_1^\mp\ell_2^\mp$ decays ($\ell = e, \mu$) can be generated by the heavy Majorana neutrino exchange, i.e by the same mechanism
that leads to the neutrinoless nuclear double $\beta$ decay. While the latter process allows to explore the first lepton generation, the $K^\pm\to\pi^\mp\ell_1^\mp\ell_2^\mp$ decays additionally provide sensitivity to the effects of Majorana neutrinos in the second generation. The searches for these processes translate into exclusion limits in the space of the parameters of the extended PMNS mixing matrix ($|U_{\ell_1 4}U_{\ell_2 4}|$) and the heavy neutrino mass $m_4$~\cite{li00, al01, at09}.

The interest in the Dirac versus Majorana nature of the neutrino stems from the fact that, for Majorana neutrinos, the see-saw mechanism provides a natural explanation for the lightness of the observed neutrino mass states. In this scenario, the existence of heavy-neutrino mass eigenstates that participate in the neutrino mixing to form sterile, right-handed neutrino flavour eigenstates is also predicted.

Until recently, the most stringent limit on the $K^\pm\to\pi^\mp\mu^\mp\mu^\mp$ decay rate came from a special data set collected in 1997 by the Brookhaven E865 $K^+$ decay-in-flight experiment~\cite{e865}. Five candidates in the signal region with an expected background from $K^+\to\pi^+\pi^+\pi^-$ decays (extrapolated from sidebands of the mass spectrum) of 5.3 events translated into an upper limit of ${\cal B}(K^+\to\pi^-\mu^+\mu^+) < 3.0\times 10^{-9}$ at 90\%~CL.

More recently, the NA48/2 experiment at CERN performed an analysis with the full $K^\pm$ decay-in-flight data sample recorded in 2003--04. Simultaneous $K^+$ and $K^-$ beams with the central momentum of 60~GeV/$c$ and a narrow momentum band entered a 114~m long vacuum decay tank, downstream of which helium-filled spectrometer consisting of four drift chambers and an analyzing magnet with a transverse momentum kick of 120 MeV/$c$ was used to track and analyze $K^\pm$ decay products. A trigger scintillator hodoscope, a liquid Krypton electromagnetic calorimeter (LKr), an iron/scintillator hadronic calorimeter, and muon detectors were located downstream of the spectrometer.

Both $K^\pm\to\pi^\mp\mu^\mp\mu^\mp$ (signal) and $K^\pm\to\pi^\pm\pi^+\pi^-$ (normalization) samples were collected with the same trigger for three-track decays. For both signal and normalization events, three-track vertices with no significant missing momentum reconstructed from the magnetic spectrometer information were required. Identification of pion and muon candidates was performed on the basis of energy deposition in the LKr calorimeter and the muon detector. The muon identification efficiency was measured to be above 98\% for momentum above 10~GeV/$c$.

\begin{figure}[t]
\begin{center}
\resizebox{0.50\textwidth}{!}{\includegraphics{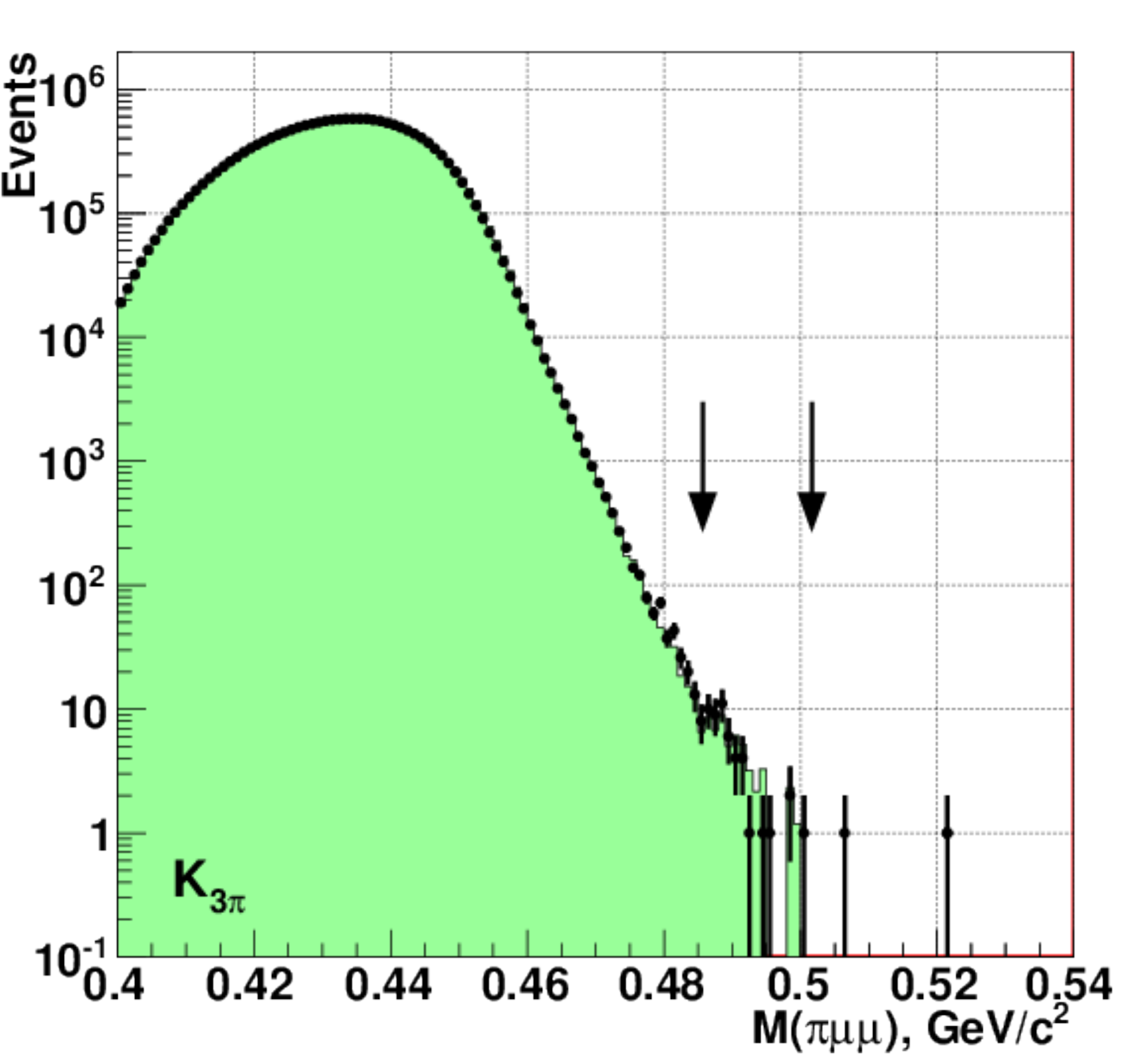}}
\end{center}
\vspace{-7mm}
\caption{Invariant mass distribution of LFV $K^\pm\to\pi^\mp\mu^\pm\mu^\pm$ candidates from the NA48/2 data sample. Signal region limits are indicated with vertical arrows.}
\label{fig:pimm}
\end{figure}

The invariant mass distribution for $K^\pm\to\pi^\mp\mu^\pm\mu^\pm$ candidates events is shown in Fig.~\ref{fig:pimm}, together with the background expectation from Monte Carlo (MC) simulations. The background is due to $K^\pm\to\pi^\pm\pi^+\pi^-$ decays, either  followed by $\pi^\pm\to\mu^\pm\nu$ decays, or with pions misidentified as muons by the LKr and muon detectors. An important circumstance contributing to pion and muon mis-identification is the finite granularity of the muon detectors; in particular, the counters are shaped as slabs rather than pads.

The 52 $K^\pm\to\pi^\mp\mu^\pm\mu^\pm$ candidates in the signal mass region with an expected background of $52.6\pm19.8$ events, where the error is systematic, translate into an upper limit ${\cal B}(K^\pm\to\pi^\mp\mu^\pm\mu^\pm) < 1.1\times10^{-9}$ at 90\% CL, conservatively assuming that the geometrical acceptance is the smallest of those for the $K^\pm\to\pi^\pm\mu^\pm\mu^\mp$ and $K^\pm\to\pi^\pm\pi^+\pi^-$ decays ($A_{\pi\mu\mu}=15.4\%$ and $A_{3\pi}=22.2\%$). This is an improvement upon the E865 result by about a factor of three. The sensitivity of this measurement is however limited by the systematic uncertainty of the background estimate, rather than the single event sensitivity. Further details of the analysis are given in Ref.~\cite{ba11}.

Being a by-product of a measurement of the allowed flavour changing neutral current decay $K^\pm\to\pi^\pm\mu^+\mu^-$, the analysis is not optimized at achieve maximum sensitivity to the LFV decay, and does not exploit the single event sensitivity of $\sim 3\times 10^{-11}$ provided by the available data sample. A dedicated re-analysis of the NA48/2 data set can potentially improve the upper limit by an order of magnitude.

\boldmath
\section{NA62 prospects for LFV in $K^+$ decays}
\unboldmath

\begin{figure*}[t]
\begin{center}
\resizebox{\textwidth}{!}{\includegraphics{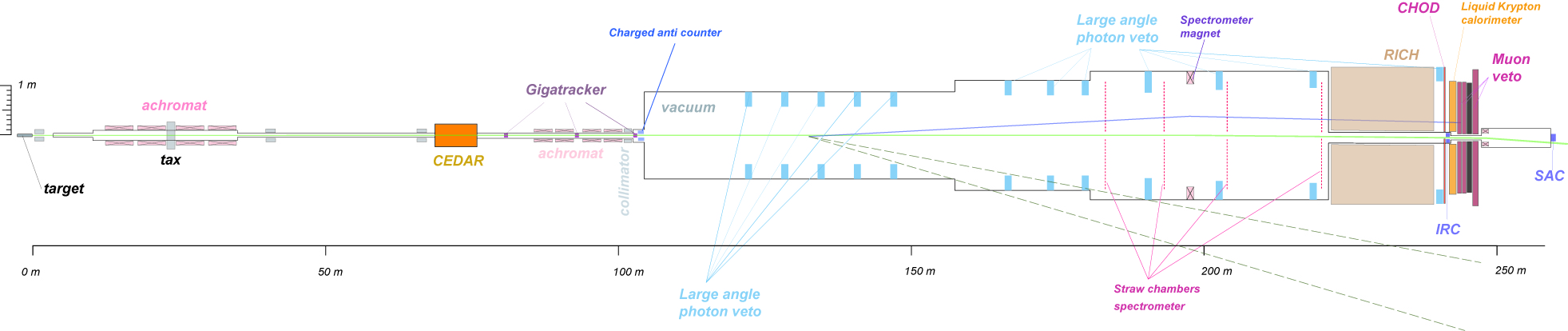}}
\end{center}
\vspace{-7mm}
\caption{Schematic diagram of the NA62 setup.}
\label{fig:na62-setup}
\end{figure*}

The NA62 experimental setup is illustrated schematically in Fig.~\ref{fig:na62-setup}.
The main subdetectors are: a differential Cherenkov counter (CEDAR) on the beam line to identify the $K^+$ in the beam and provide precise time measurements (sub-100~ps resolution); a silicon pixel beam tracker (Gigatracker); anticounters surrounding the beam tracker to veto catastrophic interactions; a downstream spectrometer composed of four straw chambers operating in vacuum; a RICH detector designed to identify pions, muons and electrons and provide precise time measurements; a scintillator hodoscope (CHOD) providing the trigger signals; a muon veto detector composed of three planes of counters. The photon veto detectors include 12 large angle annular lead glass calorimeters surrounding the decay and detector volume, the NA48 liquid Krypton (LKr) calorimeter and two small angle calorimeters (IRC, SAC), which collectively provide hermetic coverage for photons from $K^+$ decays in the fiducial vacuum decay region. Part of the experimental infrastructure, as well as the LKr calorimeter, have been inherited from the previous NA48 experiments. A more detailed description of the experimental setup and its status is given in Ref.~\cite{ferdi}.

The experiment will use an unseparated 75 GeV/$c$ positive beam with a total rate of about 800~MHz, dominated by $\pi^+$ and with a 6\% $K^+$ component. The $K^+$ decay rate inside the 65~m long fiducial volume located in the vacuum tank will amount to about 5~MHz. The expected total number of $K^+$ decays in the fiducial decay volume over the lifetime of the experiment is $1.2\times 10^{13}$ $K^{+}$, which is a 50-fold increase in statistics relative to NA48/2.

The NA62 experiment will be equipped by improved (with respect to NA48/2) redundant particle identification systems, including a RICH detector and a high granularity muon detector system, which will be crucial for background rejection. Moreover, it will achieve improved resolution on kinematic variables with respect to NA48/2. The momenta of individual beam particles with be measured (which was not the case for NA48/2), and the downstream straw spectrometer operating in vacuum with an increased spectrometer magnet momentum kick (270 MeV/$c$) and a small total thickness of $1.8\%X_0$ will provide momentum measurements for the charged secondaries with excellent resolution. As a result, mass resolutions will improve by about a factor of two with respect to NA48/2. In particular, the $K^+\to\pi^+\pi^+\pi^-$ invariant mass resolution will improve from 1.7~MeV/$c^2$ (NA48/2) to better than 1 MeV/$c^2$ (NA62). This is directly relevant for background suppression in LFV analyses, constraining the signal mass regions. For the $K^+\to\pi^-\mu^+\mu^+$ analysis, the higher momentum kick provided by the spectrometer will eliminate the tail towards higher values of the reconstructed $\pi^-\mu^+\mu^+$ invariant mass from $K^+\to\pi^+\pi^+\pi^-$ decays followed by $\pi^+\to\mu^+\nu$ decays in the spectrometer, which is a limiting factor in the NA48/2 data (Fig.~\ref{fig:pimm}).

Preliminary trigger strategy studies have been performed for the LFV decays listed in Table~\ref{tab:lfv}. These decays are characterized by three charged particles in the final state (note that a $\pi^0$ is emitted in $K^+$ decay together with a charged particle). The expected rate of a generic level-0 (L0) three-track trigger is about 0.5~MHz (dominated by $K^+\to\pi^+\pi^+\pi^-$ decays). It is comparable to the total design L0 rate of 1~MHz. Therefore the strategy of triggering on all three-track decays (adopted earlier by NA48/2) is not feasible for the high beam rate environment of NA62, and calorimetric L0 trigger conditions have to be considered additionally. The following L0
trigger primitives have been considered:
\begin{itemize}
\item $Q_N$: hits in at least $N$ hodoscope quadrants;
\item ${\rm LKr}_N(x)$: at least $N$ clusters with energy greater than $x$ GeV in the LKr calorimeter;
\item ${\rm MUV}_N$: Hits in at least $N$ pads of the fast muon detector (MUV3).
\end{itemize}
The following possible charge blind L0 di-lepton trigger conditions can be formed from these primitives: $Q_2 \times {\rm LKr}_2(15)$ for $ee$ pairs; $Q_2 \times {\rm MUV}_2$ for $\mu\mu$ pairs; $Q_2 \times {\rm LKr}_1(15) \times {\rm MUV}_1$ for $\mu e$ pairs. The total L0 trigger rate provided by these or similar conditions is expected to be below 100~kHz (mainly due to $K^+\to\pi^+\pi^+\pi^-$ decays and time coincidences of multiple $K^+$ decays and/or beam halo muons), and can be accommodated within NA62 data acquisition. The electron momentum cut-off at 15~GeV/$c$ is essential to suppress the rate generated by the $K^+\to\pi^+\pi^+\pi^-$ decays, and leads to a partial loss of acceptance for modes with $e^\pm$ in the final state.

The above trigger logic will allow the collection of a wide range of rare $K^+$ and $\pi^0$ decays in addition to the LFV signatures. Examples include $K^+\to\pi^+\ell^+\ell^-$ and $\pi^0\to e^+e^-$ decays, as well as a large $\pi^0\to\gamma e^+e^-$ sample sensitive to the dark photon production~\cite{batell09}.

Considering the expected number of $K^+$ decays in the fiducial volume and the expected acceptances (typically $\sim 10\%$), the NA62 single event sensitivities are $\sim 10^{-12}$ for $K^+$ decays and $\sim 10^{-11}$ for $\pi^0$ decays. Assuming low background levels (which is justified at least for the $K^+\to\pi^-\mu^+\mu^+$ mode, as discussed above), NA62 is well positioned to improve significantly on the current state-of-the-art reported in Table 1.

\boldmath
\section{Lepton universality tests in $K^\pm\to\ell^\pm\nu$ decays}
\unboldmath

The SM predictions for the leptonic decay rates of pseudoscalar mesons $P^\pm\to\ell^\pm\nu$ (denoted $P_{\ell 2}$ below) are affected by hadronic and CKM uncertainties. However the ratios of decay rates of the same parent meson depend only on kinematic factors and radiative corrections, and can be computed very precisely. The SM prediction for the ratio $R_K={\cal B}(K_{e2})/{\cal B}(K_{\mu2})$, inclusive of internal bremsstrahlung radiation, is $R_K = (2.477\pm0.001)\times 10^{-5}$~\cite{ci07}. The suppression of the $K_{e2}$ decay due to angular momentum conservation, as well as the high precision of the SM prediction, make the quantity $R_K$ interesting for precision searches for evidence of physics beyond the SM.

Within extensions of the SM involving two Higgs doublets, $R_K$ is sensitive to lepton flavour violating effects induced by loop processes with the charged Higgs boson exchange~\cite{ma06}. It has been argued that $R_K$ can be enhanced by ${\cal O}(1\%)$ within the Minimal Supersymmetric Standard Model~\cite{gi12}. On the other hand, the potential new physics effects are constrained by other observables such as $B^0\to\mu^+\mu^-$ and $B^+\to\tau^+\nu$ decay rates~\cite{fo12}. $R_K$ is also sensitive to the neutrino mixing parameters within SM extensions involving a fourth generation of quarks and leptons~\cite{la10} or sterile neutrinos~\cite{ab12}.

\begin{figure}[t]
\begin{center}
\resizebox{0.45\textwidth}{!}{\includegraphics{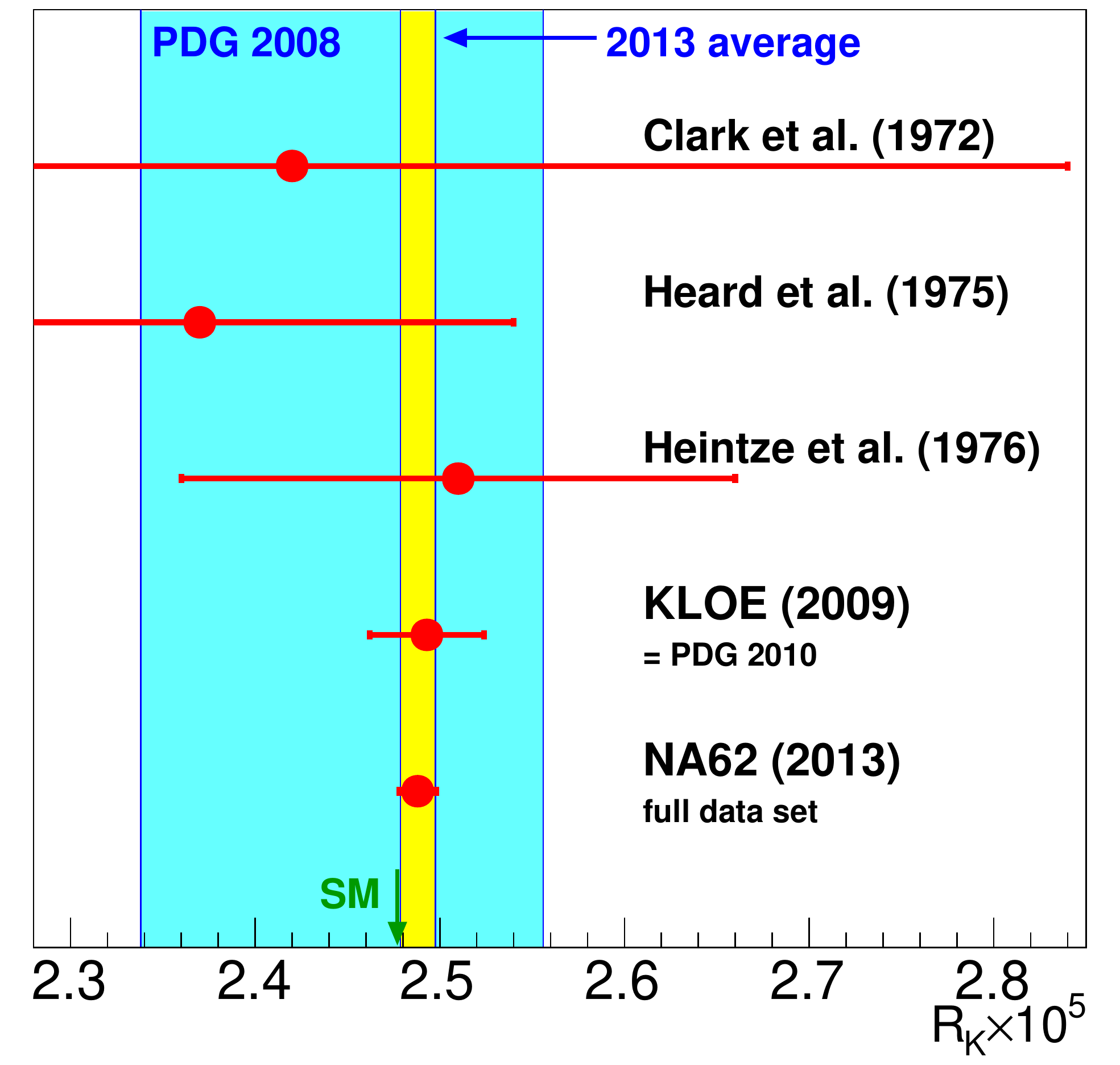}}
\end{center}
\vspace{-7mm}
\caption{Summary of $R_K={\cal B}(K_{e2})/{\cal B}(K_{\mu2})$ measurements.}
\label{fig:rk}
\end{figure}

The $R_K$ phase of the NA62 experiment (NA62-$R_K$), which took data in 2007--2008 using beam line and detector of the earlier NA48/2 experiment, has been optimized for a precision $R_K$ measurement. To minimize the systematic effects, the data were recorded at low beam intensity with a minimum bias trigger; the total number of $K^\pm$ decays during the data taking amounted to $\sim 10\%$ of that for the NA48/2 experiment. A sample of about $1.5\times 10^5$ $K^\pm_{e2}$ decays with a background of about 10\% was collected. The final result of a measurement based on that data set is $R_K=(2.488\pm0.007_{\rm stat}\pm0.007_{\rm syst})\times10^{-5}$~\cite{la13}. The leading systematic uncertainties come from background subtraction (the background is due $K^\pm$ decays and the muon halo of the kaon beams). The result is consistent with the earlier measurements and with the SM expectation and dominates the current world average, as seen in Fig.~\ref{fig:rk}.

The experimental uncertainty on $R_K$ is still an order of magnitude larger than the uncertainty on the SM prediction, which motivates further measurements at improved precision. The NA62 experiment can potentially achieve smaller uncertainties: the larger data sample will improve the statistical precision and will allow considering only the kinematic regions with lower background contamination. At the same time, improved particle identification, hermetic photon veto and the availability of detectors in the kaon beam line (CEDAR, Gigatracker) will eliminate several background sources. A sample of $\sim 10^6$ $K_{e2}$ decays can be collected by NA62 using minimum bias trigger conditions, such as $Q_1$ defined in Section 4, downscaled by a factor of $\sim 50$.

\boldmath
\section{Search for heavy neutrinos in $K^\pm\to\mu^\pm\nu$ decay}
\unboldmath

\begin{figure}[t]
\begin{center}
\resizebox{0.45\textwidth}{!}{\includegraphics{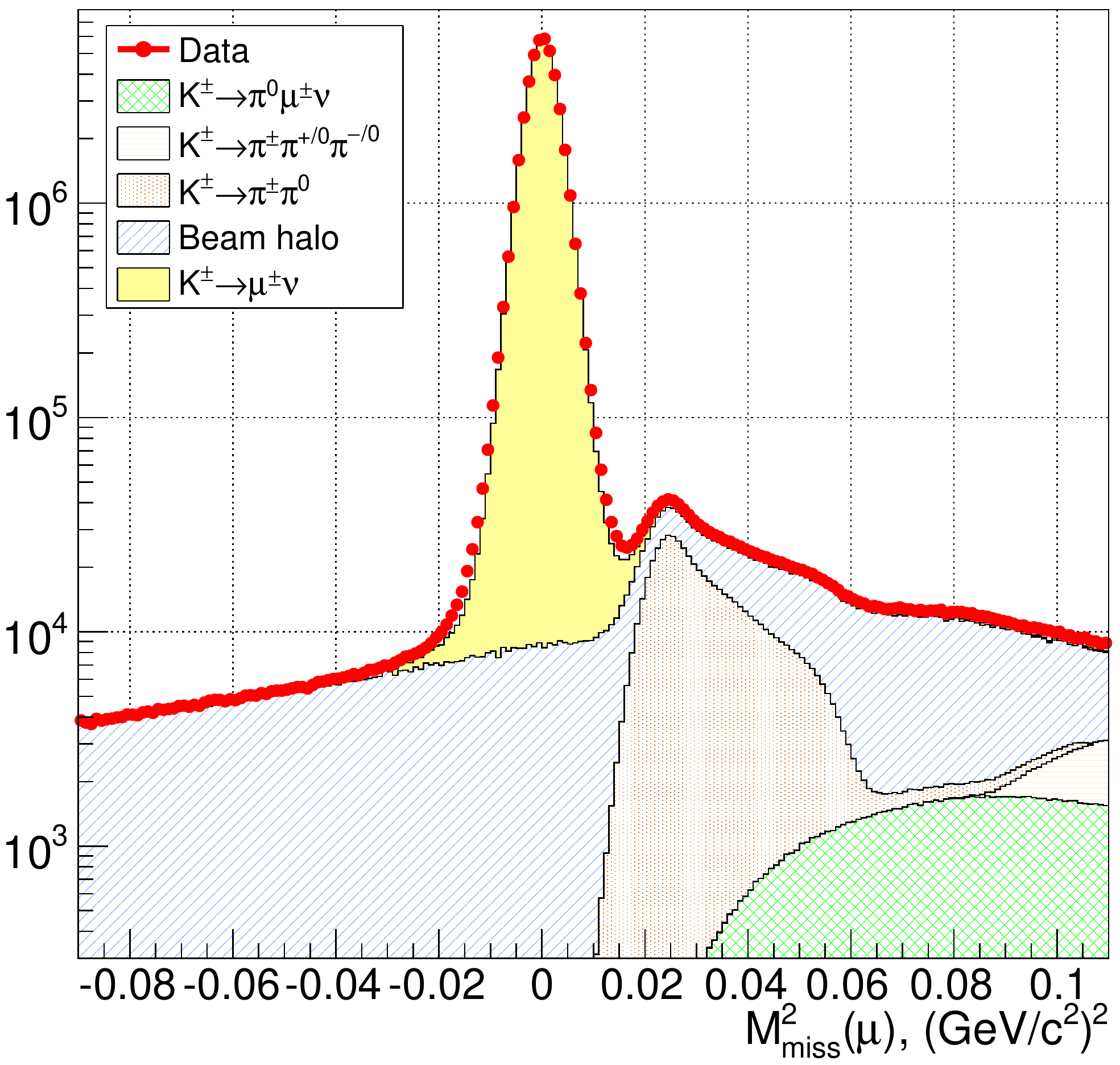}}
\end{center}
\vspace{-7mm}
\caption{Distribution of reconstructed squared missing mass $M_{\mathrm{miss}}^2(\mu)$ of the $K_{\mu 2}$ candidates, compared with the sums of normalized estimated signal and background components, for a subset (about 40\%) of the minimum data sample collected by the NA62-$R_K$ experiment in 2007--2008. Muon identification is not used for this analysis.}
\label{fig:km2}
\end{figure}

The large samples of $K^\pm\to\ell^\pm\nu$ decays collected in 2007--2008 allow precision searches for heavy neutrino mass eigenstates $\nu_H$ produced in these decays. The missing mass spectrum of the $K_{\mu2}$ decay candidates for a subset of the 2007 $K_{\mu2}$ sample used for the measurement of $R_K = {\cal B}(K_{e2})/{\cal B}(K_{\mu2})$~\cite{la13}, corresponding to $1.8\times 10^7$ $K_{\mu2}$ events, is shown in Fig.~\ref{fig:km2}. The employed ``production search'' technique is based on the search for statistically significant peaks above background in the region of positive missing mass, corresponding to two-body decays. The event signature consists of a single track positively identified as a muon and consistent with originating from a beam kaon decay, with no other activity in the detector. Unlike the alternative ``decay search'' technique looking for specific decay modes of the heavy neutrinos within detector, this method is sensitive to long-lived heavy neutrinos escaping the detector. This analysis is sensitive to $\nu_H$ masses up to $m_K-m_\mu=390~{\rm MeV}/c^2$. In practice, the lower boundary of the sensitivity region is determined by the missing mass resolution of the $K_{\mu2}$ peak, and is about 100~MeV/$c^2$ in this case.

In the absence of backgrounds, the sensitivity of the analyzed data set to heavy neutrino mixing parameters (the extended PMNS matrix element characterizing the coupling of the heavy-mass eigenstate to the $\mu$ flavour eigenstate) would be $|U_{\mu H}|^2 \sim 10^{-7}$ in the $\nu_H$ mass range of $100~{\rm MeV}/c^2< m_H< 390~{\rm MeV}/c^2$. However, as indicated by preliminary studies, the backgrounds from beam halo muons as well as hadronic and semileptonic $K^\pm$ decays limit the sensitivity to the level between $|U_{\mu H}|^2 = 10^{-6}$ and $|U_{\mu H}|^2 = 10^{-5}$ over most of the above mass range. This means that the analysis is potentially capable of placing the most stringent limits on $|U_{\mu H}|^2$ assuming a neutrino in a high-mass range ($300~{\rm MeV}/c^2 < m_H < 390~{\rm MeV}/c^2$).

As in the $R_K$ case, the NA62 experiment can improve the $|U_{\mu H}|^2$ limits significantly and is expected to have a wider $m_H$ reach, due to much smaller background level and improved mass resolution.

\section{Summary}

Next-generation experiments to measure $K^+\to\pi^+\nu\bar\nu$ will be characterized by
intense $K^+$ sources and robust background rejection, therefore they will be well positioned to carry out a rich programme to search for very rare or forbidden $K^+$ and $\pi^0$ decays.

With $\sim 10^{13}$ $K^+$ decays in its fiducial volume over the data taking, the NA62 experiment will have single event sensitivities of the order of $10^{-12}$ ($10^{-11}$) for a number of $K^+$ ($\pi^0$) decays that violate lepton flavour and/or number conservation, as well as the potential to improve existing limits in a variety of searches for related phenomena. Related measurements performed with data sets collected during the preceding stages of the kaon experimental programme at CERN serve as a proof of principle, and have already achieved record precisions.

\nocite{*}
\bibliographystyle{elsarticle-num}
\bibliography{martin}



\end{document}